\RequirePackage{fix-cm}
\documentclass[smallextended]{svjour3}       
\smartqed  
\usepackage{graphicx,amsfonts,subfigure,amsmath,amssymb}
\journalname{Circuits Syst Signal Process}
\begin{document}

\title{A Class of Deterministic Sensing Matrices and Their Application in Harmonic Detection
\thanks{This work was supported by the National Natural Science Foundation of China with grant number 61401315.}
}


\author{Shan Huang          \and
        Hong Sun \and
        Lei Yu \and
        Haijian Zhang
}


\institute{Shan Huang \\
           staronice@whu.edu.cn           
           \and \\
           Hong Sun \\
           hongsun@whu.edu.cn
           \and \\
           Lei Yu \\
           ly.wd@whu.edu.cn
           \and \\
           Haijian Zhang \\
           haijian.zhang@whu.edu.cn
           }
\date{Received: date / Accepted: date}
\maketitle
\begin{abstract}
In this paper, a class of deterministic sensing matrices are constructed by selecting rows from Fourier matrices. These matrices have better performance in sparse recovery than random partial Fourier matrices. The coherence and restricted isometry property of these matrices are given to evaluate their capacity as compressive sensing matrices. In general, compressed sensing requires random sampling in data acquisition, which is difficult to implement in hardware. By using these sensing matrices in harmonic detection, a deterministic sampling method is provided. The frequencies and amplitudes of the harmonic components are estimated from under-sampled data. The simulations show that this under-sampled method is feasible and valid in noisy environments.
\end{abstract}

\section{Introduction}
\label{intro}

Compressed sensing (CS) theory asserts that one can recover sparse signals from far fewer samples or measurements than traditional methods \cite{candes2008introduction}. The concept of CS is to recover high dimension sparse signals through low dimension measurements. The two fundamental questions in compressed sensing are how to construct suitable sensing matrices and how to recover sparse signals from under-sampled data efficiently. To find the sparsest solution, one would solve the problem
\begin{equation}\label{eq1}
  \arg {\min }{\left\| \vec x \right\|_0}\quad s.t. \quad {\vec{\Phi }}\vec x =\vec y,
\end{equation}
where $\vec x\in \mathbb{C}^{N}$ is a sparse vector, $\ell_{0}$ norm denotes its number of nonzero elements, $\vec{\Phi}\in \mathbb{C}^{{M} \times {N}}$ is called a (compressed) sensing matrix and $\vec y\in \mathbb{C}^M$ is the measurement. However, this is a hard combinatorial problem. In order to make problem (\ref{eq1}) solvable, the sensing matrix $\bf{\Phi}$ must obey a uniform uncertainty principle.

The sensing matrix $\vec{\Phi}$ has the ($\emph{k},\delta$)-\emph{restricted isometry property (RIP)} if
\begin{equation}\label{eq2}
\left( {1 - {\delta}} \right)\left\| \vec x \right\|_2^2 \le \left\| {{\vec{\Phi }}\vec x} \right\|_2^2 \le \left( {1 + {\delta}} \right)\left\| \vec x \right\|_2^2
\end{equation}
holds for all \emph{k}-sparse vectors $\vec x$, $\|\vec x\|_{2}$ denotes $\ell_{2}$-norm of $\vec x$ \cite{candes2008introduction}. The smallest $\delta$ for ($\emph{k},\delta$)-RIP is the restricted isometry constant (RIC) $\delta_{k}$. If a complex matrix $\vec{\Phi}$ satisfies (${k},\delta$)-RIP, then the $2M\times 2N$ real matrix $\vec{\Phi'}$, formed by replacing each element $a+\sqrt{-1}b$ by matrix {\small $\left( {\begin{array}{*{20}{c}}a&{ - b}\\b&a\end{array}} \right)$ } also satisfies the same (${k},\delta$)-RIP in real number domain \cite{bourgain2011breaking}. Let $\vec{\Phi}$ be a matrix with $\ell_{2}$-normalized columns $\varphi_{1},\varphi_{2},\cdots,\varphi_{N}$, i.e. $\|\varphi_{n}\|_{2}=1$ for $n=1,2,\cdots,N$, the condition (\ref{eq2}) is equivalent to that the Gram matrix $\vec{\Phi}_{\mathcal{K}}^{H}\vec{\Phi}_{\mathcal{K}}$ of every column submatrix $\vec{\Phi}_{\mathcal{K}}(\mathcal{K}\subset \{1,2,\cdots,N\}, |\mathcal{K}|\leq k)$ has all its eigenvalues in the interval $[1-\delta_{k},1+\delta_{k}]$. If $\delta_{2k}<1$, the problem (\ref{eq1}) has an unique $k$-sparse solution. It has been proven that when $\delta_{2k}<\sqrt{2}-1$ the problem (\ref{eq1}) can be approximated by a relaxed $\ell_{1}$ norm convex optimization \cite{candes2008restricted}, that is
\begin{equation}\label{eq2a}
  \arg {\min }{\left\| \vec x \right\|_1}\quad s.t. \quad {\vec{\Phi }}\vec x =\vec y.
\end{equation}

Another criterion often used to evaluate the property of a CS matrix is \emph{coherence}. For the matrix $\vec{\Phi}$ with $\ell_{2}$-normalized columns, the coherence of $\vec{\Phi}$ is defined as
\begin{equation}\label{eq3}
 \mu \left( {\vec{\Phi }} \right) = {\max _{1 \le n \ne n' \le N}}\left| {\left\langle {{\varphi _n},{\varphi _{n'}}} \right\rangle } \right|.
\end{equation}
Although the computation of coherence which only involves two columns each time is much more feasible, it considers the worst case and often leads to results somewhat pessimistic.

By the concentration inequalities, some random matrices are proven to satisfy RIP with high probability, such as matrices with independent Gaussian or Bernoulli elements \cite{baraniuk2008simple}, or matrices whose rows are randomly selected from the discrete Fourier transform matrices \cite{rudelson2008sparse}. In practical application, random sensing matrices usually imply random sampling in data acquisition, which is difficult to implement in hardware. Many scholars have become interested in finding deterministic RIP matrix constructions \cite{tang2014deterministic}\cite{zeng2015deterministic}. A deterministic construction of sensing matrices was given by polynomials over finite fields and the RIP weaker than random matrices was proven \cite{devore2007deterministic}. The class of partial Fourier matrices is of special importance in CS. Some deterministic sensing matrices of partial Fourier were constructed in \cite{haupt2010restricted} and \cite{xu2015compressed}. Sensing matrices similar to partial Fourier matrices were also constructed explicitly \cite{yu2013deterministic}, even some matrices break the $k^{2}$ barrier related to RIP \cite{bourgain2011breaking}.

In this paper, we shall verify that a class of deterministic partial Fourier matrices can be used as sensing matrices and utilize these matrices to design a deterministic sampling method for harmonic detection. It will be demonstrated that the frequencies and amplitudes of harmonic components can be estimated from deterministic sub-Nyquist samples.
\section{Deterministic Sensing Matrices}
\label{sec:1}
In this section, we give the sensing matrices directly and analyze their property as sensing matrices. Let $\mathbf{F}_{N}$ be the $N\times N$ Fourier matrix whose element in $m$-th row and $n$-th column is given by
\begin{equation}\label{eq4}
  \vec{F}_{N}[m,n]=\exp\left(\frac{j 2\pi m n}{N}\right),~m,n=1,2,\cdots,N,
\end{equation}
where $j=\sqrt{-1}$ is imaginary unit. We restrict $N=4z+3$ ($z$ is a positive integer) to be a prime number and choose $M=2z+1=(N-1)/{2}$ rows from $\vec{F}_{N}$ to construct an $M \times N$ partial Fourier matrix. The indexes of the rows are given by
\begin{equation}\label{eq5}
  \mathcal{M}=\left\{ {m:p \cdot {m^2}\bmod N,~m = 1,2, \cdots ,M} \right\},
\end{equation}
where $p$ is an arbitrary positive integer co-prime to $N$. The $M$ row indexes just don't repeat according to number theory. Because of the periodicity of trigonometric functions, the elements of generated matrix $\vec{A}$ can be written as
\begin{equation}\label{eq6}
 {\bf{A}}\left[ {m,n} \right] = \exp \left( {\frac{{j 2\pi\cdot p{m^2} \cdot n}}{N}} \right),~m\in\mathcal{M}, n=1,2,\cdots,N.
\end{equation}
\subsection{The Coherence}
Then we check the property of the matrix $\vec{A}$. In order to calculate the coherence of $\vec{A}$, we introduce the following theorem about quadratic Gauss sum:
\begin{theorem}[\rm{Theorem 1.5.2 in \cite{berndt1998gauss}}]
Let $k$ and $N$ be coprime integers with $N>0$ and $N\equiv3(\bmod 4)$.Then
\[g\left( {k;N} \right) \triangleq \sum\limits_{m = 0}^{N - 1} {\exp \left( {\frac{{j2\pi k{m^2}}}{N}} \right)}  = \left( {\frac{k}{N}} \right)j\sqrt N, \]
where $\left( {\frac{k}{N}} \right)$ denotes the Jacobi symbol.
\end{theorem}

The coherence of $\mathbf{A}$ can be expressed as
\begin{equation}\label{eq7}
  \mu \left( {\bf{A}} \right) = {\max _{d\in\{1,2\cdots,N - 1\}}}\frac{1}{M}\left| {\sum\limits_{m = 1}^M {\exp \left( {\frac{{j 2\pi dp {m^2}}}{N}} \right)} } \right|.
\end{equation}
Taking note that the values of $(dpm^{2}\bmod N)$ distribute symmetrically when $m=1,2,\cdots,N-1$, so we obtain
\begin{equation}\label{eq8}
\mu \left( \vec A \right) = \left| {\frac{{ \pm j\sqrt N  - 1}}{{2M}}} \right| = \frac{{\sqrt {M + 1} }}{{\sqrt 2 M}}.
\end{equation}

It has been proven that orthogonal matching pursuit (OMP) and basis pursuit (BP) both can recover the $ k$-sparse solution of (\ref{eq1}) when $k < \frac{1}{2}\left( {{\mu ^{ - 1}} + 1} \right)$\cite{tropp2004greed}. So $\vec{A}$ can guarantee the recovery of sparse signals below the sparse level of $\frac{1}{2}\left( {\frac{{\sqrt 2 M}}{{\sqrt {M + 1} }} + 1} \right) \approx \sqrt {\frac{M}{2}}$. As what metioned in Sect.~\ref{intro}, the result is somewhat pessimistic compared with numerical experiments.
\subsection{The RIP}
We give the RIP of $\vec{A}$ according to the following result:
\begin{theorem}[\rm{Theorem 14.1 in \cite{foucart2013mathematical}}]
Suppose $\vec x$ is a $k$-sparse vector which is uniformly distributed among all $k$-sparse vectors. If for $\delta,\varepsilon\in(0,1)$, there exists a constant $c>0$ such that
\begin{equation}\label{eq8a}
   \mu \left( {\bf{\Phi }} \right) \le \frac{{c\delta }}{{\ln \left( {{N \mathord{\left/
 {\vphantom {N \varepsilon }} \right.
 \kern-\nulldelimiterspace} \varepsilon }} \right)}},\frac{k}{N}\left\| {\bf{\Phi }} \right\|_2^2 \le \frac{{c{\delta ^2}}}{{\ln \left( {{N \mathord{\left/
 {\vphantom {N \varepsilon }} \right.
 \kern-\nulldelimiterspace} \varepsilon }} \right)}}.
\end{equation}
Then $\vec{\Phi}$ with $\ell_{2}$-normalized columns satisfies ($\emph{k},\delta$)-RIP with probability at least $1-\varepsilon$.
\end{theorem}

\begin{figure}[!htbp]
\centering
\subfigure[]{
\includegraphics[scale=0.95]{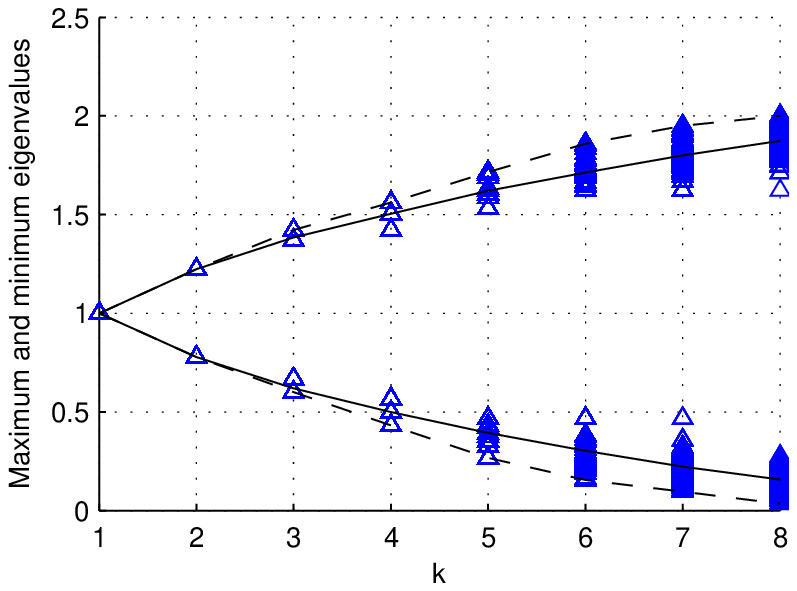}}
\subfigure[]{
\includegraphics[scale=0.95]{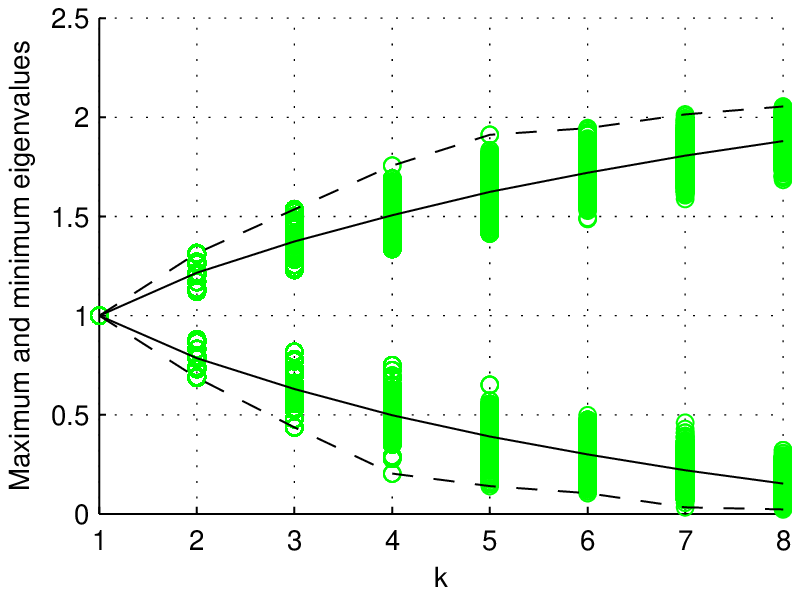}}
\caption{ Maximum and minimum eigenvalues of sub-Gram matrices for $k$. (a) $\vec{A}$; (b) The random partial Fourier matrix. } \label{f1}
\end{figure}

To present the statistical RIP of $\mathbf{A}$ intuitively, we plot the maximum and minimum eigenvalues of Gram matrices $\vec{\Phi}_{\mathcal{K}}^{H}\vec{\Phi}_{\mathcal{K}}$, where $\vec{\Phi}_{\mathcal{K}}$ are randomly selected column submatrices. We set $M=11$ and $N=23$. The Maximum and minimum eigenvalues of sub-Gram matrices $\vec{\Phi}_{\mathcal{K}}^{H}\vec{\Phi}_{\mathcal{K}}$ are equivalent to $1+\delta$ and $1-\delta$. These eigenvalues of a random partial Fourier matrix are also plotted for comparison. The data are obtained from $2000$ sub-Gram matrices for each $k$. The solid lines sketch the average values of maximum and minimum eigenvalues of all sub-Gram matrices and dashed lines sketch the limiting values. Fig.~\ref{f1} shows that the eigenvalues of $\vec{A}$'s sub-Gram matrices distribute more intensively than the random partial Fourier matrix. In fact, the correlation of any two columns of $\vec{A}$ is either $(j\sqrt N  - 1)/2$ or $(-j\sqrt N  - 1)/2$ with equal probability, which makes the variants of sub-Gram matrices far fewer.

\begin{figure}[!htbp]
\centering
\includegraphics[scale=0.95]{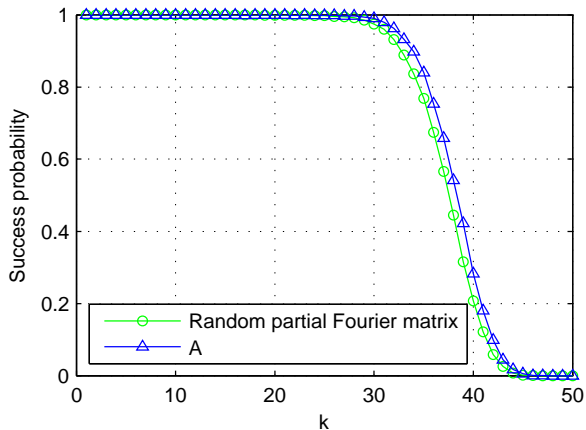}
\caption{ Comparison of recovery performance between $\mathbf{A}$ and random partial Fourier matrices. } \label{f2}
\end{figure}

Then we plot the success probabilities of recovering $k$-sparse vectors $\vec x$ via $\vec{A}$ and random partial Fourier matrices when $M=51$ and $N=103$. For convenience we use OMP algorithm. At every time, the locations of nonzero elements in vector $\vec x$ distribute uniformly among all $k$-sparse vectors and the values of nonzero elements follow standard normal distribution. If the locations of nonzero elements in $\vec x$ can be found by OMP, we say that this trial succeeds. There are $10000$ trials for every $k$. Fig.~\ref{f2} shows that $\vec{A}$ has better performance than random partial Fourier matrices.
\section{The Application in Harmonic Detection}
\label{sec:2}
Harmonic detection is widely used in electrical power systems \cite{chen2012detection}\cite{lin2007intelligent}\cite{zhang2009precise}. In \cite{lev2000application}, a staggered under-sampling algorithm is proposed to detect the harmonic. The algorithm reorders sub-Nyquist samples before applying the fast Fourier transform (FFT). We shall demonstrate that even fewer samples are enough by using CS. In this section, we analyze the problem and derive our under-sampled method in complex number domain. Then the practical implementation in real number domain is given later. The following discussion is in noiseless situations, but the method is also robust to noise as simulations show.
\subsection{Problem Formulation}
The signal $s(t)$ contains $K$ harmonic frequency components and we intend to estimate the frequencies and corresponding amplitudes. The samples can be expressed as the sums of $K$ complex exponentials, namely
\begin{equation}\label{eq9}
s(t) = \sum\limits_{k = 1}^K {{C_k}\exp \left( {j\left( {2\pi {f_k} \cdot t + {\theta _k}} \right)} \right)},
\end{equation}
where $C_{k},f_{k}$ and $\theta_{k}$ represent the amplitude, frequency and phase angle of the $k$-th harmonic component, respectively. The frequencies are all integer multiples of some fundamental frequency $f_0$, i.e. $f_{k}\in \{f_0,2f_0,\cdots,Nf_0\}$.

Suppose we have an analog-digital converter with sampling rate $f_{S}=1/\Delta t$, arbitrary samples at time points $\Delta t,2\Delta t,\cdots,l\Delta t,\cdots$ can be obtained. In general, the sampling rate $f_{S}$ must be higher than the Nyquist rate. But because the harmonic signal is sparse in frequency domain, we can use a specific low-rate analog-to-digital converter to detect harmonic components.
\subsection{The Method for Under-sampling}
Since the harmonic components of the signal are aligned with a discrete frequency grid, the sample $s[l]$ at time point $l\Delta t$ can be written as
\begin{equation}\label{eq10}
s\left[ l \right] = \sum\limits_{n = 1}^N {{{\tilde C}_n}\exp \left( {j\left( {2\pi {{\tilde f}_n} \cdot l\Delta t + {{\tilde \theta }_n}} \right)} \right)}.
\end{equation}
Certainly, without regard to noise, $\tilde{C}_{n}$ corresponding to $\tilde{f}_{n}=nf_0$ $(n=1,2,\cdots,N)$ mostly equals zero except the frequency components really contained in $s(t)$.

We give the constraint conditions which should be satisfied and the data we need directly. Denoting $p=\tilde{f}_{N}/f_{S}$ and $M=(N-1)/2$, we only need to sample at rate $f_{S}=1/\Delta t$ to obtain $M$ samples $s_1,s_2,\cdots,s_M$ at time points $\mathcal{L}$, provided that the parameters satisfy
\begin{itemize}
\item $N$ is a prime number and $N=4z+3 (z\in \mathbb{Z}^{+})$;
\item $p$ is an arbitrary positive integer coprime to $N$;
\item $\mathcal{L}=\{ l\cdot \Delta t:l=m^{2} \bmod N, m=1,2,\cdots,M\}$.
\end{itemize}
For instance, if $N=11,M=5$, we start to sample at the time $t=0$, then we should get the samples at the time points $1\Delta t, 3\Delta t, 4\Delta t, 5\Delta t, 9\Delta t$. Noticing that $p=\tilde{f}_{N}/f_{S}$ is simply required to be coprime to $N$, the sampling rate $f_{S}$ may be much lower than the Nyquist rate. Generally speaking, $N$ does not always meet the first condition, we can select another satisfactory $N'\geq N$ and a proper sampling rate.

Then the detection method is explained as follows. Because
\begin{equation}\label{eq11}
 \exp \left( {j2\pi {{\tilde f}_n}l\Delta t} \right) = \exp \left( {j2\pi \frac{{{{\tilde f}_n} \cdot l \cdot p}}{{{\tilde{f}_N}}}} \right)
 = \exp \left( {\frac{{j2\pi  \cdot pl \cdot n}}{N}} \right),
\end{equation}
applying the third condition above, we have
\begin{equation}\label{eq12}
{s_m} = \sum\limits_{n = 1}^N {{{\tilde C}_n}\exp \left( {j\left( {\frac{{2\pi  \cdot p{m^2} \cdot n}}{N} + {{\tilde \theta }_n}} \right)} \right)} ,
\end{equation}
where $m = 1,2, \cdots ,M$. We let
\begin{equation}\label{eq13}
\vec y = \left( {\begin{array}{*{20}{c}}
{s_1}\\
{s_2}\\
 \vdots \\
{s_M}
\end{array}} \right),
\vec x = \left( {\begin{array}{*{20}{c}}
{{{\tilde C}_1}{e^{j{{\tilde \theta }_1}}}}\\
{{{\tilde C}_2}{e^{j{{\tilde \theta }_2}}}}\\
 \vdots \\
{{{\tilde C}_N}{e^{j{{\tilde \theta }_N}}}}
\end{array}} \right).
\end{equation}
Then
\begin{equation}\label{eq14}
\vec y = {\vec{A}}\vec x,
\end{equation}
where
\begin{equation}\label{eq15}
{\vec{A}} = \left( {\begin{array}{*{20}{c}}
{{\exp(\frac{{j2\pi  \cdot p{1^2} \cdot 1}}{N})}}&{{\exp(\frac{{j2\pi  \cdot p{1^2} \cdot 2}}{N})}}& \cdots &{\exp(\frac{{j2\pi  \cdot p{1^2} \cdot N}}{N})}\\
{{\exp(\frac{{j2\pi  \cdot p{2^2} \cdot 1}}{N})}}&{{\exp(\frac{{j2\pi  \cdot p{2^2} \cdot 2}}{N})}}& \cdots &{{\exp(\frac{{j2\pi  \cdot p{2^2} \cdot N}}{N})}}\\
 \vdots & \vdots & \ddots & \vdots \\
{{\exp(\frac{{j2\pi  \cdot p{M^2} \cdot 1}}{N})}}&{{\exp(\frac{{j2\pi  \cdot p{M^2} \cdot 2}}{N})}}& \cdots &{{\exp(\frac{{j2\pi  \cdot p{M^2} \cdot N}}{N})}}
\end{array}} \right).
\end{equation}
In (\ref{eq14}), $\vec y$ and $\vec{A}$ are known. If $\vec x$ can be solved from (\ref{eq14}), $\tilde{C}_{n}$ and $\tilde{\theta}_{n}$ corresponding to $\tilde{f}_{n}$ will be known. Obviously, $\vec x$ is a $K$-sparse vector. This is the same type of problem with (\ref{eq1}). The coherence and RIP of $\vec{A}$ have been verified in Sect.~\ref{sec:1}, so the problem (\ref{eq14}) can be solved by common sparse recovery algorithms. That is to say, after sampling in accordance with the above requirements, the frequencies and amplitudes of the harmonic components can be calculated.

\subsection{Practical Implementation}
In practical applications, real numbers are handled instead of complex numbers. We give the implementation of our method in real number domain. For the convenience of the following discussion, we assume that the last element of $\vec x$ is 0, so the last column of $\vec{A}$ can be abandoned. Then (\ref{eq11}) changes into
\begin{equation}\label{eq16}
\vec y = \left( {\begin{array}{*{20}{c}}
{\vec{B}}&{{{\vec{B}}^ * }}
\end{array}} \right)\left( {\begin{array}{*{20}{c}}
{{\vec x_{1}}}\\
{{{{\bar{\vec x}}}_{{2}}}}
\end{array}} \right),
\end{equation}
where $\vec {B}$ consists of the first $M$ columns of $\vec{A}$ and $\vec{B}^{*}$ is its conjugate, $\vec x_{1}$ consists of the first $M$ elements of $\vec x$ and $\bar {\vec x}_2$ consists of the second $M$ elements of $\vec x$ adversely. By separating the real part and imaginary part, (\ref{eq16}) turns into
\begin{equation}\label{eq17}
\begin{split}
 {\vec y_r} + j{\vec y_i} &= {{\vec{B}}_r}{\vec x_{1r}} - {{\vec{B}}_i}{\vec x_{1i}} + {{\vec{B}}_r}{{\bar{\vec x}}_{2r}} + {{\vec{B}}_i}{{\bar{\vec x}}_{2i}}\\
 &+j\left( {{{\vec{B}}_i}{\vec x_{1r}}} + {{\vec{B}}_r}{\vec x_{1i}} - {{\vec{B}}_i}{{\bar{\vec x}}_{2r}} + {{\vec{B}}_r}{{\bar{\vec x}}_{2i}} \right),
\end{split}
\end{equation}
that is,
\begin{equation}\label{eq18}
\vec y' = {\vec{B'}}\vec x',
\end{equation}
where
\begin{equation}\label{eq19}
\vec y' = \left( {\begin{array}{*{20}{c}}
{{\vec y_r}}\\
{{\vec y_i}}
\end{array}} \right),
\vec x' = \left( {\begin{array}{*{20}{c}}
{{\vec x_{1r}}}\\
{{\vec x_{1i}}}\\
{{\bar{\vec x}_{2r}}}\\
{{\bar{\vec x}_{2i}}}
\end{array}} \right),
{\vec{B'}} = \left( {\begin{array}{*{20}{c}}
{{{\vec{B}}_r}}&{ - {{\vec{B}}_i}}&{{{\vec{B}}_r}}&{{{\vec{B}}_i}}\\
{{{\vec{B}}_i}}&{{{\vec{B}}_r}}&{ - {{\vec{B}}_i}}&{{{\vec{B}}_r}}
\end{array}} \right).
\end{equation}
The subscripts $r$ and $i$ denote the real part and imaginary part of a matrix respectively. The compound matrix $\vec{B}'$ in (\ref{eq18}) satisfies the same $(k,\delta)$-RIP with $\vec{A}$ in real number domain. However, in practical measurement, such as in power systems, we couldn't get the imaginary part $\vec y_{i}$. We solve this problem by assuming that $\bar{\vec x}_{2}$ of the actual signal is $\vec 0$. That is to say, the highest harmonic frequency which can be detected is reduced by half. Comparing with the theoretical analysis in complex number domain, we should choose twice the highest frequency as upper limit in practice.

In the process of recovering $\vec x'$, we set $\bar{\vec x}_{2r}=\vec x_{1r}$ and $\bar{\vec x}_{2i}=-\vec x_{1i}$, namely
\begin{equation}\label{eq20}
\vec y' = \left( {\begin{array}{*{20}{c}}
{2{\vec y_r}}\\ \vec 0
\end{array}} \right),
\vec x' = \left( {\begin{array}{*{20}{c}}
{\vec x_{1r}}\\
{\vec x_{1i}}\\
{\vec x_{1r}}\\
 - {\vec x_{1i}}
\end{array}} \right).
\end{equation}
By using CS recovery algorithms, we obtain $\vec x_{1r}$ and $\vec x_{1i}$. Then the amplitude $\tilde{C}_{n}$ and phase angle $\tilde{\theta}_{n}$ with respect to the discrete frequency $\tilde{f}_{n} (n=1,2,\cdots,M)$ can be determined.
\section{Simulation Results }
In this section, we shall create satisfactory signals and estimate their spectrums by the proposed method. The signals are composed of several cosine waves. The frequency components are assumed to distribute randomly in $\{1Hz,2Hz,\cdots,50Hz\}$. The amplitudes and phase angles of the frequency components are set to be uniformly distributed in $[0.1,1]$ and $[0,2\pi)$. According to our method, we set $N=103$ and $M=51$. The sampling frequency $f_{S}$ is set to be $\tilde{f}_{N}/100=1.03Hz$. The number of nonzero frequency components is known as the sparsity $k$.

\begin{figure}[!htbp]
\centering
\includegraphics[scale=0.95]{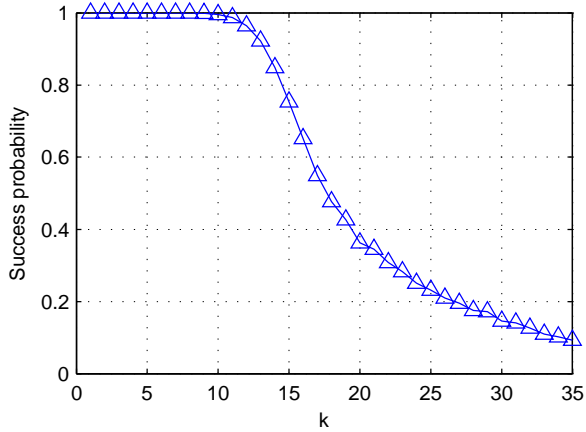}
\caption{ Success probability of estimation depending on sparsity in noiseless condition. } \label{f3}
\end{figure}

Firstly the situation without noise is considered. We use OMP as before to recover the signals and count the probability of success for varying $k$. The probability of success for every $k$ is the result of 10000 trials. In Fig.~\ref{f3}, we can see the success probability is 1 when $k\leq10$, which is much smaller than the critical value shown in Fig.~\ref{f2}. This is because the actual sparsity of $\vec x'$ in solving process is no longer $k$. In fact, the locations of nonzero elements of $\vec x'$ have strong correlation because $\vec x'$ consists of a complex vector's real part and imaginary part. If the recovery algorithm is modified to take advantage of this structural sparsity, better performance will be achieved.

\begin{figure}[!htbp]
\centering
\includegraphics[scale=0.95]{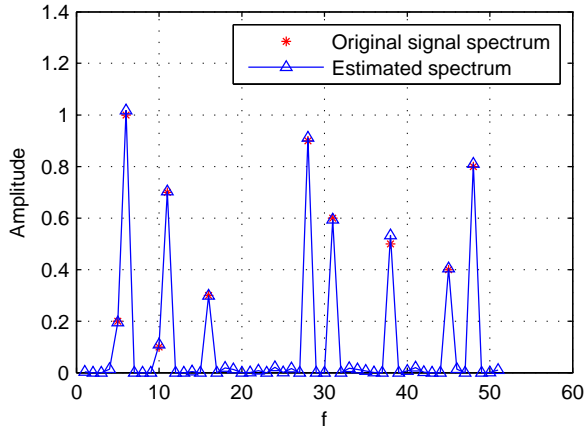}
\caption{ Comparison between original signal spectrum and estimated spectrum. } \label{f4}
\end{figure}

Then we estimate the spectrum of a frequency-sparse signal in white Gaussian noise with SNR=30dB. The signal has 10 frequency components and the corresponding amplitudes are $0.1,0.2,\cdots,0.9,1$ respectively. The estimated spectrum is compared with original signal as shown in Fig.~\ref{f4}. The frequency components are all found with little error.

\begin{figure}[!htbp]
\centering
\includegraphics[scale=0.95]{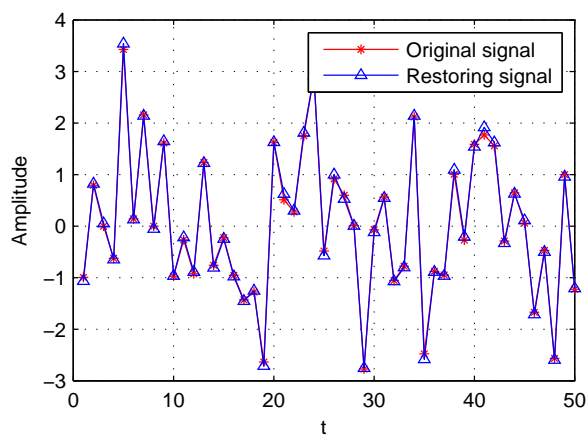}
\caption{ Comparison between original signal and restoring signal in time domain. } \label{f5}
\end{figure}

At last we reconstruct the signal according to the amplitudes and phase angles we solve, even though it is not our aim. The reconstructed signal is plotted to compare with the original signal in Fig.~\ref{f5}. The reconstructed signal and the original signal are both sampled at $100Hz$ in time domain. The relative accumulated error of 50 points is about $3.85\%$.
This indicates that the samples at the Nyquist rate can be obtained approximatively from under-sampled data.

In above simulations, the sampling frequency has no influence on the results, provided that $p=\tilde{f}_{N}/f_{S}$ is coprime to $N$. The errors decrease as the sparsity $k$ decreases and the SNR increases, which is the same as standard CS model.
\section{Conclusion}
In this paper, we construct a class of deterministic sensing matrices and verify their property as sensing matrices. Based on the special structure of the matrices, a deterministic under-sampled method of harmonic detection is proposed. Very few samples are required to estimate the frequencies and amplitudes of the harmonic components. Through theoretical analysis and numerical experiments, the feasibility and robustness of the method have been verified.


\bibliographystyle{spmpsci}      
\bibliography{mybibfile}   


\end{document}